\documentclass[%
reprint,
superscriptaddress,
showpacs,showkeys,
 amsmath,amssymb,
 aps,
floatfix,
]{revtex4-1}

\usepackage{graphicx}
\graphicspath{{Figures/}}
\usepackage{dcolumn}
\usepackage{bm}
\usepackage[table]{xcolor}
\usepackage[colorlinks=true,citecolor=blue,linkcolor=red,urlcolor=violet]{hyperref}

\makeatletter

    \def\CT@@do@color{%
      \global\let\CT@do@color\relax
            \@tempdima\wd\z@
            \advance\@tempdima\@tempdimb
            \advance\@tempdima\@tempdimc
    \advance\@tempdimb\tabcolsep
    \advance\@tempdimc\tabcolsep
    \advance\@tempdima2\tabcolsep
            \kern-\@tempdimb
            \leaders\vrule
                    \hskip\@tempdima\@plus  1fill
            \kern-\@tempdimc
            \hskip-\wd\z@ \@plus -1fill }
    \makeatother
\newcommand{\blue}{\color{blue}}

\begin{document}

\preprint{Surf. Interfaces}

\title{Epitaxial growth and characterization of (001) [NiFe/M]$_{20}$ (M = Cu, CuPt and Pt) superlattices}

\author{Movaffaq Kateb}
 \email{movaffaq.kateb@chalmers.se}
\affiliation{Science Institute, University of Iceland, Dunhaga 3, IS-107 Reykjavik, Iceland}
\affiliation{Condensed Matter and Materials Theory Division, Department of Physics, Chalmers University of Technology, SE-412~96 Gothenburg, Sweden}
\author{Jon Tomas Gudmundsson}
 \email{tumi@hi.is}
\affiliation{Science Institute, University of Iceland, Dunhaga 3, IS-107 Reykjavik, Iceland}
\affiliation{Space and Plasma Physics, School of Electrical Engineering and Computer Science, KTH Royal Institute of Technology, SE-100~44, Stockholm, Sweden}
\author{Snorri Ingvarsson}%
 \email{sthi@hi.is}
\affiliation{Science Institute, University of Iceland, Dunhaga 3, IS-107 Reykjavik, Iceland}

\date{\today}

\begin{abstract}
We present optimization of [(15\,{\AA})~Ni$_{80}$Fe$_{20}$/(5\,{\AA})~M]$_{20}$ single crystal multilayers on (001) MgO, with M being Cu, Cu$_{50}$Pt$_{50}$ and Pt. These superlattices were characterized by high resolution X-ray reflectivity (XRR) and diffraction (XRD) as well as polar mapping of important crystal planes. It is shown that cube on cube epitaxial relationship can be obtained when depositing at substrate temperature of 100~$^\circ$C regardless of the lattice mismatch (5\% and 14\% for Cu and Pt, respectively). At lower substrate  temperatures poly-crystalline multilayers were obtained while at higher  substrate temperatures \{111\} planes appear at $\sim$10$^\circ$ off normal to the film plane. It is also shown that as the epitaxial strain increases, the easy magnetization axis rotates towards the direction that previously was assumed to be harder, i.e.\ from [110] to [100], and eventually further increase in the strain makes the magnetic hysteresis loops isotropic in the film plane. Higher epitaxial strain is also accompanied with increased coercivity values. Thus, the effect of epitaxial strain on the magnetocrystalline anisotropy is much larger than what was observed previously in similar, but polycrystalline samples with uniaxial anisotropy (Kateb \textit{et al.} 2021).
\end{abstract}

\maketitle

\section{Introduction}
Since the discovery of the giant magneto-resistance (GMR) effect by Fert \citep{baibich1988} and Gr{\"u}nberg \citep{binasch1989} in the late 1980s, magnetic multilayers have been widely studied. In many cases they present unique features that cannot be achieved within the bulk state namely inter-layer exchange coupling \citep{parkin90:2304}, magnetic damping, due to the interface \citep{mizukami2001, ingvarsson2002} rather than alloying \citep{ingvarsson2004}, and perpendicular magnetic anisotropy \citep{johnson1996}. 

The GMR discovery, without a doubt, was an outcome of the advances in preparation methods such as molecular beam epitaxy (MBE), that enabled deposition of multilayer films with nanoscale thicknesses \citep{fert2008}. Thus, a great deal of effort has been devoted to enhancing the preparation methods over the years using both simulations \citep{zhou1998,zhou2004a,zhou2004,ene05:3383} and experiments (cf.\ Ref.~\citep{kateb21:168288} and references therein). Permalloy (Py) multilayers with non-magnetic (NM) Pt \citep{correa16:115,kateb21:168288,leydominguez15:17D910} or Cu \citep{rook91:5670,vonloyen00:4852,hecker02:62,ene05:3383,heitmann00:4849}  as spacers have been studied extensively in recent years.  Various deposition methods have been utilized for preparing magnetic multilayers such as MBE \citep{rook91:5670},  pulsed laser deposition (PLD) \citep{shen2004}, ion beam deposition \citep{ueno95:2194,ene05:3383}, dc magnetron sputtering (dcMS) \citep{parkin90:2304,correa16:115,hecker02:62,vonloyen00:4852}, and more recently,  high power impulse magnetron sputtering (HiPIMS) \citep{kateb21:168288}.

Permalloy (Py) is a unique material with regards to studying magnetic anisotropy, which has been shown to strongly depend on the preparation method \citep{kateb2019epi}. For instance, uniaxial anisotropy can be induced in polycrystalline Py by several means \citep{kateb2021}. However, it has been thought that the cubic symmetry of single crystal Py encourages magneto-crystalline anisotropy, while uniaxial anisotropy cannot be achieved. We have recently shown that using HiPIMS deposition one can decrease the Ni$_3$Fe (L1$_2$) order, but maintain the single crystal form, to achieve uniaxial anisotropy. We attributed this to the high instantaneous deposition rate during the HiPIMS pulse \citep{kateb2021md}, which limits ordering compared to dcMS that present cubic (biaxial) anisotropy. Regarding Py multilayers there has been a lot of focus on magneto-dynamic properties recently while the effects of interface strain on magnetic anisotropy has not received much attention. \citet{rook91:5670} prepared polycrystalline Py/Cu multilayers by MBE and reported a weak anisotropy in them, i.e.\ hysteresis loops along both the hard and easy axes with complete saturation at higher fields. They compared the coercivity values ($H_{\rm c}$) and saturation fields of their samples to $H_{\rm c}$ and anisotropy field ($H_{\rm k}$) of sputter deposited multilayers showing uniaxial anisotropy and concluded that the latter gives more than twice harder properties. They also reported an increase in $H_{\rm c}$ with Py thickness and attributed this to the interface strain that relaxes with increased thickness. \citet{correa16:115} prepared nanocrystalline Py/Pt multilayers on rigid and flexible substrates and in both cases obtained weak anisotropy but two orders of magnitude larger $H_{\rm c}$. Unfortunately, they did not mention any change in magnetic anisotropy upon straining the flexible substrate. 

Recently we showed that utilizing increased power to the dcMS process, and in particular, by using HiPIMS deposition that the interface sharpness in polycrystalline [Py/Pt]$_{20}$ multilayers can be improved, due to increased ionization of the sputtered species \citep{kateb21:168288}.
{\blue Briefly,} in dcMS deposition the film forming material is composed mostly of neutral atoms \citep{gudmundsson20:113001}, while in HiPIMS deposition a significant fraction of the film forming material consists of ions \citep{gudmundsson12:030801,lundin20b}. In fact we have shown that higher ionization of the film-forming material leads to smoother film surfaces and sharper interfaces using molecular dynamics simulations \cite{kateb2019md,kateb2020md}.  
We also showed that by changing the non-magnetic spacer material one can increase interface strain that is accompanied with higher $H_{\rm c}$, $H_{\rm k}$ and limited deterioration of uniaxial anisotropy \cite{kateb21:168288}.

Another aspect of preparation is that deposition chambers for multilayers mostly benefit from oblique deposition geometry, which encourage uniaxial anisotropy in Py. The origin of uniaxial anisotropy induced by oblique deposition has been proposed to be self-shadowing, but this has not been systematically verified. We demonstrated uniaxial anisotropy, even in atomically smooth films with normal texture, which indicates lack of self-shadowing \citep{kateb2017,kateb2019}. We also showed that oblique deposition is more decisive in definition of anisotropy direction than application of an \textit{in-situ} magnetic field for inducing uniaxial magnetic anisotropy. 
Also for polycrystalline Py films oblique deposition by HiPIMS presents a lower coercivity and anisotropy field than when dcMS deposition is applied \citep{kateb2018,hajihoseini2019,kateb2019phd}. While none of the above mentioned results verify self-shadowing they are consistent with our interpretation of the order i.e.\ oblique deposition induces more disorder than \textit{in-situ} magnetic field and HiPIMS produce more disorder than dcMS. Note that the level of order in polycrystals cannot be easily observed by X-ray diffraction. In this regard we proposed a method for mapping the resistivity tensor that is very sensitive to level of order in Py \citep{ingvarsson2017,kateb2021}.
We reported much higher coercivity and deterioration of uniaxial anisotropy in (111)\,Py/Pt multilayers obtained by HiPIMS deposition of the Py layers \citep{kateb21:168288}. We attributed the latter effect to the interface sharpness and higher epitaxial strain when HiPIMS is utilized for Py deposition. 

Here, we study the properties of Py superlattices deposited by dcMS with Pt, Cu and CuPt as non-magnetic spacers. Pt and Cu were chosen as spacer because they have lattice parameters of 3.9 and 3.5\,{\AA}, respectively, and therefore provide varying strain to the Py film which has lattice constant of 3.54\,{\AA}. In this regard, calibration of the substrate temperature during deposition with respect to the desired thickness is of prime importance \cite{loloee02:4541}. It is worth mentioning that dcMS deposition is expected to give more ordered single crystal (001)~Py layers in which crystalline anisotropy is dominant \citep{kateb2019epi}. This enables understanding to what extent interface strain will affect magnetocrystalline anisotropy of Py which we will show is much larger than the changes in uniaxial anisotropy in our latest study \citep{kateb21:168288}. 
Section \ref{experimental} discusses the deposition method and process parameters for  the fabrication of the superlattices and the characterization methods applied.  In Section \ref{resultsanddiscussion} the effects of substrate temperature on the properties of the Py/Cu system are studied followed by exploring the influence of varying the lattice parameter of the non-magnetic layer on the structural and magnetic properties of the superlattice.  The findings are summarized in Section \ref{summary}.

\section{Experimental apparatus and methods}
\label{experimental}

The substrates were one side polished single crystal (001)\,MgO (Crystal GmbH) with surface roughness $<$5\,{\AA} and of dimensions 10\,mm$\times$10\,mm$\times$0.5\,mm. The MgO substrates were used as received without any cleaning but were baked for an hour at 600\,$^\circ$C in vacuum for dehydration, cooled down for about an hour, and then maintained at the desired  temperature $\pm$0.4\,$^\circ$C during the deposition. The superlattices were deposited in a custom built UHV magnetron sputter chamber with a base pressure below $5 \times 10^{-7}$\,Pa. The chamber is designed to support 5 magnetron assemblies and targets, which are all located 22\,cm away from substrate holder with a 35$^\circ$ angle with respect to substrate normal. The shutters were controlled by a LabVIEW program (National Instruments). The deposition was made with argon of 99.999\,\% purity as the working gas using a Ni$_{80}$Fe$_{20}$ at.\% and Cu targets both of 75\,mm diameter and a Pt target of 50\,mm in diameter.

The Py depositions were performed at 150\,W dc power (MDX 500 power supply from Advanced Energy) at argon working gas pressure of 0.25\,Pa which gives deposition rate of  1.5\,{\AA}/s. Both pure Cu and Pt buffer layers were deposited at dc power of 20\,W. For the deposition of CuPt alloy we calibrated Cu$_{50}$Pt$_{50}$ at.\,\%  at dc power at 10 and 13\,W for Cu and Pt,  respectively. This selection of powers provide a similar deposition rate of 0.45\,{\AA}/s in all cases. In order to ensure that the film thickness is as uniform as possible, we rotate the sample at $\sim$12.8\,rpm.  These deposition processes were repeated to fabricate superlattices consisting of  20 repetitions of 15\,{\AA} Py and 5\,{\AA} Pt, Cu or Cu$_{50}$Pt$_{50}$ at.\,\% (CuPt).

X-ray diffraction measurements (XRD) were carried out using a X'pert PRO PANalitical diffractometer (Cu K$_{\alpha1}$ and K$_{\alpha2}$ lines, wavelength 0.15406 and 0.15444\,nm, respectively) mounted with a hybrid monochromator/mirror on the incident side and a 0.27$^\circ$ collimator on the diffracted side. We would like to remark that, K$_{\alpha2}$ separation at $2\theta=55^\circ$ is only 0.2$^\circ$ and much less at the smaller angles i.e. where our multilayer peaks are located. This is an order of magnitude smaller than the full width half maximum (FWHM) of our multiplayer and satellite peaks. A line focus was used with a beam width of approximately 1\,mm. The film thickness, mass density, and surface roughness, was determined by low-angle X-ray reflectivity (XRR) measurements with an angular resolution of 0.005$^\circ$, 
obtained by fitting the XRR data using the commercial X'pert reflectivity program, that is based on the Parrat formalism \citep{parratt54:359} for reflectivity.

The magnetic hysteresis was recorded using a home-made high sensitivity magneto-optic Kerr effect (MOKE) looper. We use a linearly polarized He-Ne laser of wavelength 632.8~nm as a light source, with Glan-Thompson polarizers to further polarize and to analyze the light after Kerr rotation upon reflection off the sample surface. The Glan-Thompson polarizers linearly polarize the light with a high extinction ratio. They are cross polarized near extintion, i.e.~their polarization states are near perpendicular and any change in polarization caused by the the Kerr rotation at a sample's surface is detected as a change in power of light passing through the analyzer. The coercivity was read directly from the easy axis loops. The anisotropy field is obtained by extrapolating the linear low field trace along the hard axis direction to the saturation magnetization level, a method commonly used when dealing with effective easy axis anisotropy.

\section{Results and discussion}
\label{resultsanddiscussion}

\subsection{Effect of substrate temperature on structural and magnetic properties \label{CuCu}}

\begin{figure}
    \centering
    \includegraphics[width=1\linewidth]{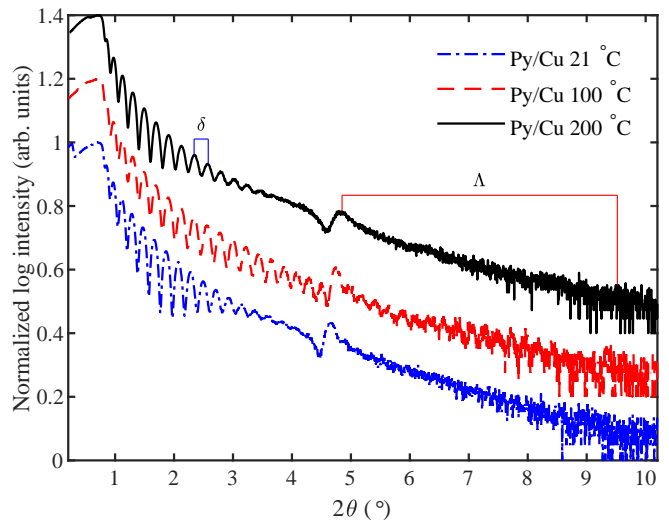}
    \caption{Comparison of the XRR pattern from [Py/Cu]$_{20}$ superlattices deposited on (001)\,MgO at different substrate temperatures. The $\Lambda$ and $\delta$ are inversely proportional to the Py/Cu period and total thickness, respectively.}
    \label{fig:XRRCuPyTsub}
\end{figure}

Figure\,\ref{fig:XRRCuPyTsub} shows the XRR results from Py/Cu superlattices deposited at different substrate temperatures. The $\Lambda$ and $\delta$ indicated in the figure are inversely proportional to the superlattice period and the total thickness, respectively. It can be clearly seen that the fringes decay faster for a Py/Cu superlattice deposited at substrate temperature of 21\,$^\circ$C and 200\,$^\circ$C than when deposited at  100\,$^\circ$C. This indicates lower surface roughness obtained in the Py/Cu superlattice deposited at 100\,$^\circ$C. When deposited at room temperature, the large lattice mismatch between MgO and Py/Cu does not allow depositing a high quality superlattice. For substrate temperature of 200\,$^\circ$C, however, it is difficult to grow a continuous Cu layer with such a low thickness (5\,{\AA}). This is due to the dewetting phenomenon which causes the minimum Cu thickness that is required to maintain its continuity to be 12\,{\AA}. Earlier, it has been shown that for substrate temperature up to 100~$^\circ$C Py/(1~{\AA}) Cu showed a limited intermixing upon annealing \citep{vonloyen00:4852}. The optimum substrate temperature for deposition obtained here is very close to 156\,$^\circ$C which has earlier been reported for the deposition of (001)\,Fe/MgO \citep{moubah2016} and (001)\,Fe$_{84}$Cu$_{16}$/MgO \citep{warnatz2020} superlattices. 
We would like to remark that in our previous study we deposited 5\,nm Ta underlayer to reduce the substrate surface roughness \citep{kateb21:168288}. However, Ta on MgO is non-trivial due to the large lattice mismatch (22\%). Besides, Ta underlyer
 encourages polycrystalline $\langle111\rangle$ texture normal to substrate surface that does not serve our purpose here.

\begin{figure}[hbt!]
    \centering
    \includegraphics[width=1\linewidth]{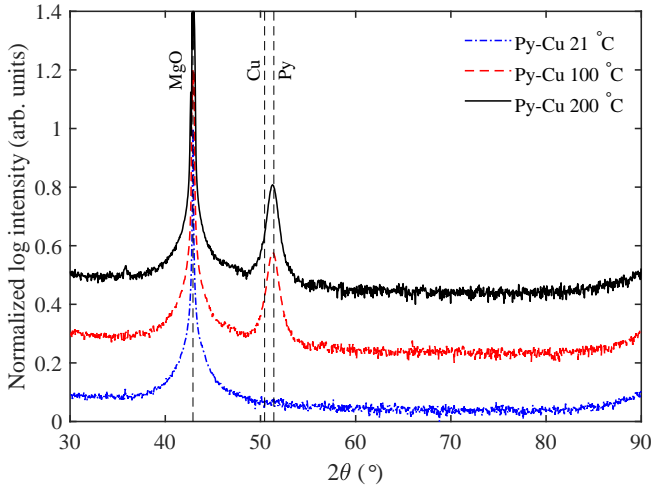}
    \caption{Comparison of the XRD pattern from [Py/Cu]$_{20}$ superlattices deposited on (001)\,MgO at different substrate temperatures. The intense peak belongs to (002) planes of MgO and and the other peak is due to (002) planes of Py/Cu multilayer.}
    \label{fig:XRDCuPyTsub}
\end{figure}

Figure\,\ref{fig:XRDCuPyTsub} shows the result of symmetric ($\theta-2\theta$) XRD scan normal to the film for Py/Cu superlattices deposited at different substrate temperatures. It can be seen that no Cu and Py peak were detected in the superlattice deposited at room temperature. Thus, epitaxial growth of Py and Cu were suppressed by the low substrate temperature. Furthermore, we studied room temperature deposited Py/Cu using grazing incidence XRD which indicated a polycrystalline structure (not shown here). For substrate temperature of 100 -- 200\,$^\circ$C there are clear (002)\,Py/Cu peaks indicating an epitaxial relationship in the (001)\,Py\,$\|$\,(001)\,Cu\,$\|$\,(001)\,MgO stack. However, there is no sign of satellite peaks due to the $\Lambda$ (Py/Cu) period. We explain this further when  comparing the Py/Cu, Py/Pt, and Py/CuPt superlattices in Section \ref{PtCuPt}.

\begin{figure}[hbt!]
    \centering
    \includegraphics[width=1\linewidth]{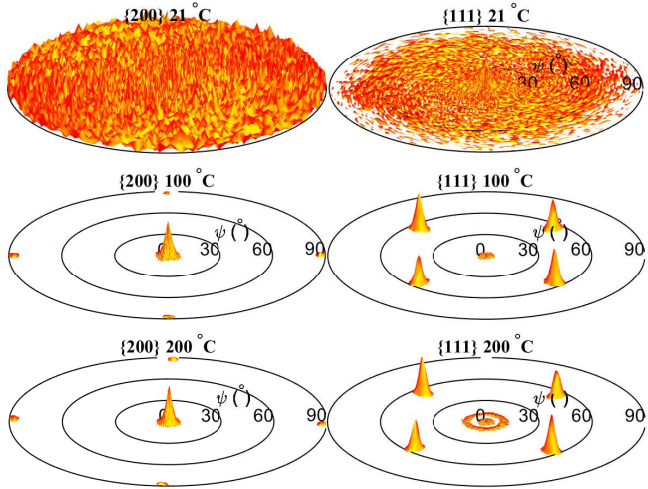}
    \caption{Comparison of the \{200\} and \{111\} pole figures from [Py/Cu]$_{20}$ superlattices deposited on (001)\,MgO at substrate temperatures of   21, 100, and 200$^\circ$C. The background is removed for better illustration.}
    \label{fig:PoleCuPyTsub}
\end{figure}

Figure\,\ref{fig:PoleCuPyTsub} shows the pole figures from the  \{200\} and \{111\} planes for Py/Cu superlattices deposited at different substrate temperatures. For the Py/Cu superlattice deposited at 21\,$^\circ$C, there is only a peak in the middle of the \{111\} pole figure that indicates a weak $\langle111\rangle$ contribution normal to the film plane. For a superlattice deposited with substrate temperature of  100\,$^\circ$C the \{200\} pole figure indicates an intense spot at $\psi=0$ that is corresponding to (002)\,Py/Cu planes parallel to the substrate. There is also a weaker four-fold  spot at $\psi=90^\circ$ and $\phi=0,90,180$ and $270^\circ$ from the \{200\} planes parallel to the substrate edges. In the \{111\} pole figure only four-fold points appear at $\psi=54.7^\circ$ and with 45$^\circ$ shifts in $\phi$ with respect to substrate edges. These are the characteristics of the \textit{so-called} cube on cube epitaxy achieved at 100\,$^\circ$C. 
For deposition with substrate temperature of  200\,$^\circ$C, however, there is a weak \{111\} ring at $\psi=7.5^\circ$. Note that these \{111\} planes were not detected by normal XRD because the (111)\,Py/Cu peak appears at $2\theta\simeq42^\circ$ which is masked by the strong (002)\,MgO peak normal to the film plane.

\begin{figure}[hbt!]
    \centering
    \includegraphics[width=1\linewidth]{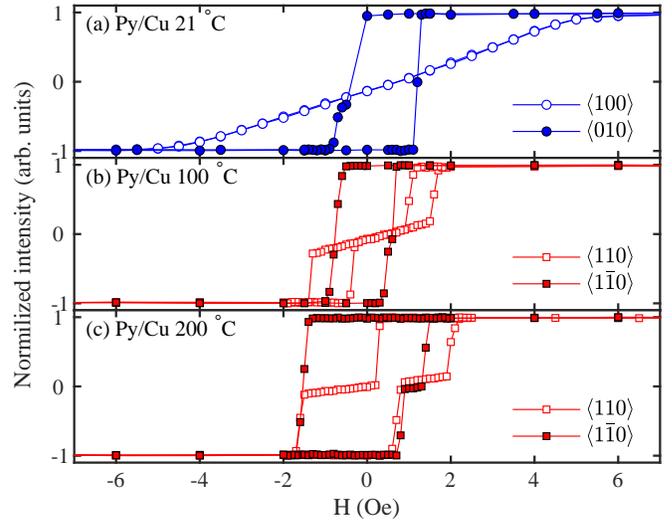}
    \caption{MOKE response of different [Py/Cu]$_{20}$ superlattices deposited on (001)\,MgO at substrate temperatures (a) 21, (b) 100 and (c) 200\,$^\circ$C. Each legend indicates probing orientation in the substrate plane.}
    \label{fig:MOKECuPyTsub}
\end{figure}

Figure\,\ref{fig:MOKECuPyTsub} compares the MOKE response of Py/Cu superlattices deposited at different substrate temperatures. For a superlattice deposited at room temperature, uniaxial anisotropy along the [100] direction is evident. This is expected since the oblique deposition in a co-deposition chamber tends to induce uniaxial anisotropy in Py films  \citep{kateb2017,kateb2018,kateb2019,kateb2019epi}. However, the oblique deposition cannot overcome magnetocrystalline anisotropy due to symmetry in an ordered single crystal Py \citep{kateb2019epi}. Thus, the low substrate temperature must be accounted for the limiting order in the Py layer and presence of uniaxial anisotropy. 

For deposition at higher substrate temperatures, however, biaxial anisotropy was obtained with the easy axes along the [110] directions in plane. It is worth mentioning that the bulk crystal symmetry gives the easy axis along the [111] direction which is forced into the film plane along the [110] direction due to shape anisotropy \citep{kateb2019epi}. In the Py/Cu superlattice grown at 100\,$^\circ$C  (Figure \ref{fig:MOKECuPyTsub} (b)), $\langle1\bar{1}0\rangle$ is clearly an easy direction, with a very low $H_{\rm c}$ of 0.7\,Oe and double-hysteresis loops along the $\langle110\rangle$ direction that saturates at 1.2\,Oe. For the Py/Cu superlattice deposited at  200$^\circ$C  (Figure \ref{fig:MOKECuPyTsub} (c)), it seems the double-hysteresis loops  overlap and the other easy axis gives a step that in total present increased coercivity. With increasing substrate temperature not only do the coercivities vary but also the shapes of the hysteresis curves are different.  When the substrate temperature during deposition  is 21$^\circ$C the magnetization, shown in Figure~\ref{fig:MOKECuPyTsub}~(a), is much like we obtain for polycrystalline single layer films. When the substrate temperature is higher, as shown in Figures~\ref{fig:MOKECuPyTsub} (b) and (c), however, the anisotropy has rotated by 45 degrees and the hysteresis loops have changed. The intermediate steps in the hysteresis curves are caused by antiferromagnetic alignment of the magnetic layers, that minimizes the exchange and dipolar magnetic interactions. In some cases this results in perfectly zero magnetic remanence, while in other cases the cancellation is not perfect. The non-magnetic Cu spacer layer is only 5~\AA \ in our case, just at the onset of the first antiferromagnetic exchange coupling peak observed by \citet{parkin1992apl}. 
Double hysteresis curves have been observed in the Py/Cu system \citep{hecker02:62}.
Note that Ni is miscible in Cu and during annealing a mixing of Ni and Cu is possible.   Such intermixing causes a decrease of magnetic homogenity and  a reduction in the GMR \citep{vonloyen00:4852,hecker02:62}.  


\subsection{Effect of strain  on structural and magnetic properties \label{PtCuPt}}

\begin{figure}[hbt!]
    \centering
    \includegraphics[width=1\linewidth]{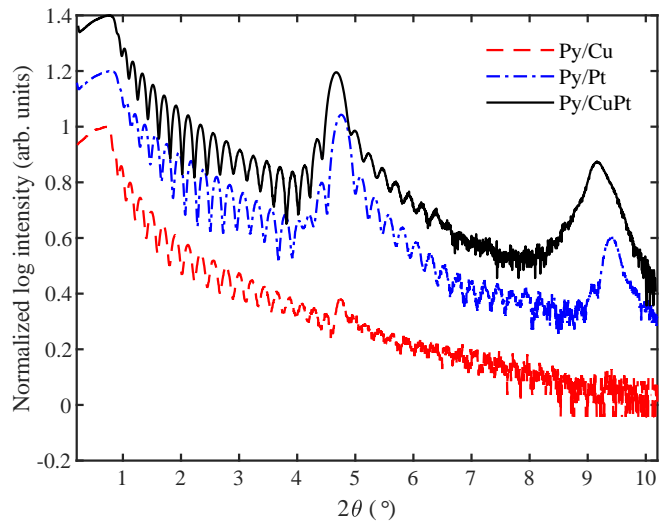}
    \caption{XRR measurements from the various  superlattices, [Py/Cu]$_{20}$, [Py/Pt]$_{20}$, and [Py/CuPt]$_{20}$, deposited on (001)\,MgO at  substrate temperature of 100\,$^\circ$C.}
    \label{fig:XRRepi}
\end{figure}

In order to explore the influence of strain on the magnetic properties we deposited NM layers of Pt and Cu$_{50}$Pt$_{50}$ at.~\% alloy in addition to Cu discussed in Section \ref{CuCu}.  Pt has lattice constant of 3.9 \AA  \  which is larger than of Py, which has lattice constant of 3.54 \AA. Therefore, by going from Cu to  Cu$_{50}$Pt$_{50}$ and then to Pt the strain is gradually increased.        
Figure\,\ref{fig:XRRepi} shows XRR results from different superlattices deposited on (001)\,MgO at 100\,$^\circ$C. Note that the $\Lambda$ peak is suppressed in the Py/Cu superlattice. One may think this arises from a diffused Py/Cu interface that leads to smooth density variation at the interface. This is not the case here and the $\Lambda$ peaks intensity decreases due the similar density of Py and Cu. The latter has been shown to reduce the resolution of the XRR measurement in Si/SiO$_{2}$ by a few orders of magnitude \cite{tiilikainen07:7497}. 

The layers thickness and their mass density as well as surface and interface roughness obtained by fitting XRR results for deposition  at substrate temperature of 100$^\circ$C  are summarized in table\,\ref{tab:xrrfit}. 
The period $\Lambda$ is in all cases about 19 \AA \  with $t_{\rm Py} \sim 16$ \AA \ and $t_{\rm NM} \sim 3$ \AA.    The film mass density of the Py layers is the highest (8.74 g/cm$^3$) in the Py/Cu stack but is lowest  (7.45 g/cm$^3$) in the Py/CuPt stack. 
\setlength{\tabcolsep}{4.2pt}
\renewcommand{\arraystretch}{1.2}
{\rowcolors{3}{gray!40}{gray!0}
\begin{table}[hbt!]
  \centering
  \caption{The Py and NM layer thicknesses ($t$), roughness (Ra) and density ($\rho$) extracted by fitting the XRR results of different superlattices deposited on (001)\,MgO at 100$^\circ$C substrate temperature.}
  \label{tab:xrrfit}
  \begin{tabular}{c|c c c | c c | c c}
    \toprule
    Sample & \multicolumn{3}{c |}{$t$ ({\AA})}  & \multicolumn{2}{c |}{Ra ({\AA})} & \multicolumn{2}{c}{$\rho$ (g/cm$^3$)} \\
    & Py & NM & $\Lambda$ & Py & NM & Py & NM \\
    \hline
    Py/Cu & 15.8 & 3.46 & 19.3 & 7.62 & 6.25 & 8.74 & 9.8 \\
    Py/Pt & 15.9 & 2.97 & 18.9 & 5.92 & 3.24 & 8.55 & 27.2 \\
    Py/CuPt & 15.9 & 3.43 & 19.3 & 2.25 & 4.94 & 7.45 & 26.8 \\
    \botrule
  \end{tabular}
\end{table}
}

\begin{figure}[hbt!]
    \centering
    \includegraphics[width=1\linewidth]{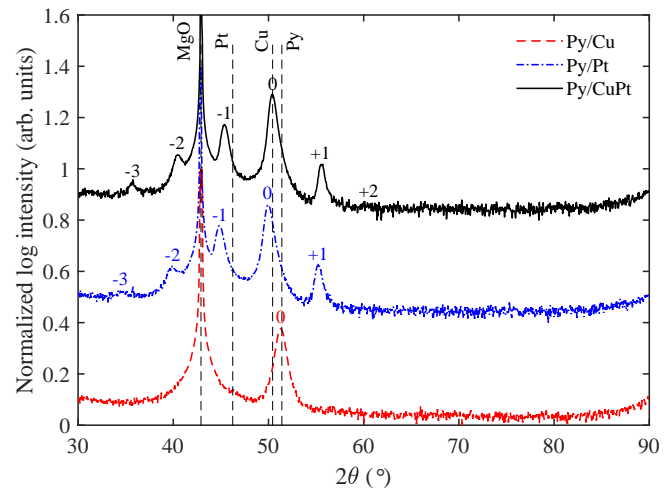}
    \caption{Comparison of the XRD results from the various  superlattices, [Py/Cu]$_{20}$, [Py/Pt]$_{20}$ and [Py/CuPt]$_{20}$, deposited at substrate temperature of 100\,$^\circ$C. All the peaks are due to the (002) plane and the vertical dashed line indicate the (002) peak position for the bulk state.}
    \label{fig:XRDepi}
\end{figure}

Figure\,\ref{fig:XRDepi} shows the XRD results from Py/NM superlattices deposited on (001)\,MgO at substrate temperature of  100\,$^\circ$C. The most intense peak, indicated by the vertical dashed line, belong to the (002) planes of the MgO substrate. Rather than exhibiting  separate peaks for Py and NM, a single main (002)\,Py/NM peak is evident from all the superlattices and indicated by $0$ in the figure. This peak is closer to the Py side due to the higher thickness of Py layers compared to the NM layers. The other peaks, indicated by $\pm$ are satellite peaks. The asymmetric intensity of the satellite peaks is associated with the strain in direction normal to the substrate. It is also clear that the main (002)\,Py/CuPt peak is located between  the Py/Cu and Py/Pt peaks. Due to the lack of any other peaks, we conclude that these superlattices are single crystalline with their $\langle001\rangle$ orientation normal to the substrate surface.
The satellite peaks suggest the period $\Lambda$ to be of 18.8 and 19.2\,{\AA} for Py/Pt and Py/CuPt, respectively, which is slightly off compared to the values given in table\,\ref{tab:xrrfit}.

\begin{figure}[hbt!]
    \centering
    \includegraphics[width=1\linewidth]{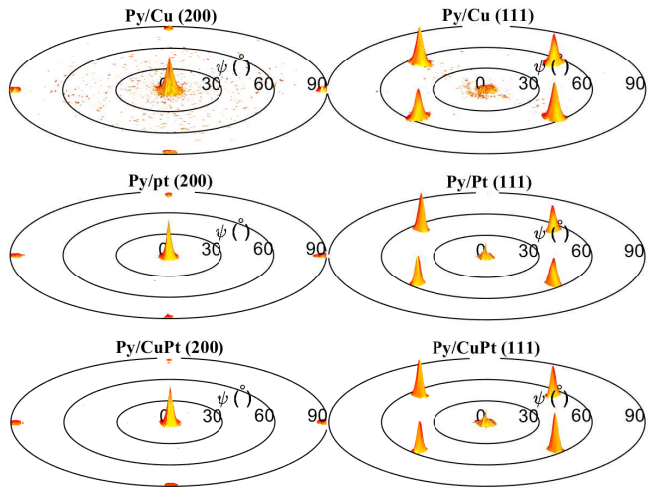}
    \caption{Comparison of the \{002\} and \{111\} pole figures from the various  superlattices, [Py/Cu]$_{20}$, [Py/Pt]$_{20}$ and [Py/CuPt]$_{20}$, deposited on (001)\,MgO at substrate temperature of 100\,$^\circ$C. The background is removed for better illustration. }
    \label{fig:Poleepi}
\end{figure}

Figure\,\ref{fig:Poleepi} shows the \{002\} and \{111\} pole figures from different superlattices deposited on (001)\,MgO at  substrate temperature of 100\,$^\circ$C. Since the Py/Pt and Py/CuPt superlattices exhibit  multiple (002) peaks the pole figure were obtained for the main peak (indicated by $0$ in figure\,\ref{fig:XRDepi}). It can be seen that all the pole figures are very similar. All these pole figures indicate a cube on cube epitaxial relationship. 

\begin{figure}[hbt!]
    \centering
    \includegraphics[width=1\linewidth]{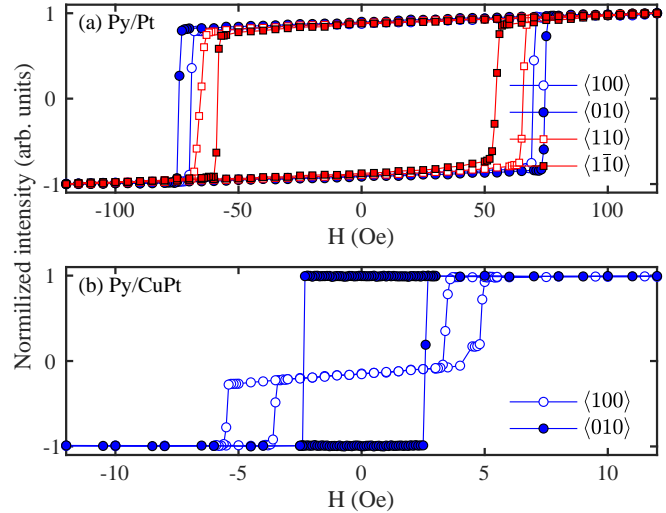}
    \caption{MOKE response from (a) [Py/Pt]$_{20}$ and (b) [Py/CuPt]$_{20}$ superlattices. The legend indicate probing direction in the substrate plane.}
    \label{fig:MOKEepi}
\end{figure}

Figure\,\ref{fig:MOKEepi} depicts the MOKE response from Py/Pt and Py/CuPt superlattices prepared at 100\,$^\circ$C. For the Py/Pt superlattice we did not detect any clear easy direction in the film plane, the film appeared almost isotropic in the film plane with $H_{\rm c}$ of 60 -- 75\,Oe. The hysteresis in the [100] and [110] directions are displayed in figure\,\ref{fig:MOKEepi}(a). Aside from $H_{\rm c}$ of 3\,Oe, the Py/CuPt superlattice presents biaxial anisotropy similar to Py/Cu, cf.\ figure\,\ref{fig:MOKECuPyTsub}(b) and (c). However, a Py/Cu superlattice exhibits an easy axis along the [110] directions, while an easy axis appears along the [100] orientations for a Py/CuPt superlattice.
Note that the [100] directions are harder than both the [110] and [111] directions. However, forcing easy axes along the [100] direction in the single crystal Py on (001)\,MgO has been reported previously \citep{michelini02:7337}. 

For the polycrystalline but highly (111) textured Py/M multilayers we observed limited change in coercivity and opening in hard axis with interface strain due to the choice of M \citep{kateb21:168288}. Here, for Py/Pt the coercivity increases an order of magnitude and cubic anisotropy is almost destroyed.

\section{Summary}
\label{summary}

In summary it is shown that Py superlattices can be successfully deposited on (001)\,MgO within a narrow substrate temperature window around 100\,$^\circ$C. For small lattice mismatch of 5\% superlattice the easy axes detected along the [110] directions is similar to the single crystal Py. It is also shown that the moderate lattice mismatch (7\%) rotates the easy axes towards the [100] orientation and the coercivity increases. The higher lattice mismatch of 14\% present nearly isotropic behaviour and a very high coercivity, simultaneously. Thus, the results indicate that the changes in magnetocrystalline anisotropy due to epitaxial strain are much larger  than the changes we observed earlier in the case of uniaxial anisotropy.  

\begin{acknowledgments}
The authors would like to acknowledge helpful comments and experimental help from Dr.~Fridrik Magnus and Einar B.\ Thorsteinsson. This work was partially supported by the Icelandic Research Fund Grant Nos.~228951, 196141, 130029 and 120002023.
\end{acknowledgments}

\bibliography{Ref,heim78}
\end{document}